\documentclass[journal]{IEEEtran}
\usepackage{cite}

\ifCLASSINFOpdf
  \usepackage[pdftex]{graphicx}
  \graphicspath{{./graphics/}}
  \DeclareGraphicsExtensions{.pdf,.jpeg,.png}
\else
\fi
\usepackage{array}

\ifCLASSOPTIONcompsoc
 \usepackage[caption=false,font=normalsize,labelfont=sf,textfont=sf]{subfig}
\else
 \usepackage[caption=false,font=footnotesize]{subfig}
\fi
\usepackage{url}

\begin{document}
%
\title{Keyshuffling Attack for Persistent Early Code Execution in the Nintendo
3DS Secure Bootchain}
%
%
%

\author{Matthew~McClintic,
        Devon~Maloney, 
        Michael~Scires, 
        Gabriel~Marcano, 
        Matthew~Norman, 
        and Aurora~Wright \vspace{-2.0em}}

%
%

\maketitle

\begin{abstract}
We demonstrate an attack on the secure bootchain of the Nintendo 3DS in order to
gain early code execution. The attack utilizes the block shuffling vulnerability
of the ECB cipher mode to rearrange keys in the Nintendo 3DS's encrypted
keystore. Because the shuffled keys will deterministically decrypt the encrypted
firmware binary to incorrect plaintext data and execute it, and because the
device's memory contents are kept between hard reboots, it is possible to
reliably reach a branching instruction to a payload in memory. This payload, due
to its execution by a privileged processor and its early execution, is able to
extract the hash of hardware secrets necessary to decrypt the device's encrypted
keystore and set up a persistent exploit of the system.
\end{abstract}

\begin{IEEEkeywords}
Advanced Encryption Standard, keyshuffling, bootchain, cryptography, block
ciphers, software security.
\end{IEEEkeywords}

\section{Introduction}

The Nintendo 3DS, like all entertainment consoles, is in a difficult position
when it comes to designing a secure system. The device must easily accommodate
legitimate users while at the same time preventing cheating, protecting
intellectual property, and enforcing system integrity. To accomplish this, the
3DS has a chain of trust based on two separate processors: an ARM9 and an ARM11.
The ARM9 processor is a security processor that runs a single process
(``Process9'') whose sole responsibility is to handle secure functions such as
cryptography, filesystem access, and permissions. The ARM11 processor is an
application processor which is responsible for all OS and userspace-level
tasks \cite{cryptosystem}.

As with most embedded systems, the root of trust for the 3DS is the boot ROM
burned into the System-on-Chip (``SoC'') at the factory. The code in this Read
Only Memory cannot be changed and contains a public key to which only Nintendo
has the matching private key. When the device is powered on, each CPU's boot ROM
loads their respective firmware binary from NAND flash storage to memory, checks
the firmware binary for a valid RSA signature that matches the burned in public
key, then jumps to the firmware binary entrypoint \cite{cryptosystem}. This is a
simple, robust chain of trust that seems fairly secure upon initial inspection.

In 2014, Nintendo released the ``New 3DS'' which this paper will focus on. This
updated 3DS features a faster CPU, more RAM, and (most importantly) an extra
encryption layer on the ARM9 firmware binary known in the 3DS community as
``ARM9Loader''. This encryption layer loads new keys from NAND sector 0x96 (the
plaintext of which is the same for all New 3DS devices), which are encrypted
with AES-128-ECB by a key calculated from a SHA-256 hash of the the
console-unique (different for every console) one-time programmable (``OTP'')
memory region of the device. Keys on the 3DS are loaded into write-only
``keyslots'', which are secure memory areas readable only by the hardware AES
implementation. This means that it should not be possible to recover these keys
after they have been written to the AES module \cite{AES_Registers}.

After ARM9Loader decrypts the NAND ``keysector'', access to the OTP memory region
is disabled until next boot via the hardware register CFG\_SYSPROT9. Once the
OTP region has been secured, ARM9Loader then decrypts the ARM9 firmware binary
using a key from the decrypted keysector. Additionally, the hash of the OTP
memory region is outputted to the SHA\_HASH hardware register after the hardware
SHA implementation calculates it \cite{SHA_Registers}. Importantly, this
register is not cleared until the ARM9 firmware binary clears it after
ARM9Loader jumps to its entrypoint.

\section{Secure Bootchain Implementations}

\subsection{Implementation (v1.0)}

The New 3DS shipped with version 8.1.0 of the system software, which contains
the following implementation of ARM9Loader in the boot process \cite{FIRM}:

\medskip
\begin{enumerate}
  \item Calculate SHA-256 hash of the OTP memory region and output the hash to
  the SHA\_HASH register
  \item Calculate AES write-only keyslot 0x11 from the OTP hash
  \item Read the keysector from NAND to memory
  \item Decrypt the keysector using keyslot 0x11
  \item Clear AES write-only keyslot 0x11 to zero
  \item Disable access to the OTP memory region by setting CFG\_SYSPROT9
  \item Write Key \#1 from the keysector to AES write-only keyslot 0x11
  \item Instruct the AES module to calculate sub-keys 0x18 through 0x1F based on
  keyslot 0x11
  \item Verify keyslot 0x11 by encrypting a fixed test vector and checking the
  result
  \item Instruct the AES module to decrypt the ARM9 firmware binary
  \item Jump to the ARM9 firmware binary entrypoint
\end{enumerate}
\medskip

The problem with this implementation of ARM9Loader is that keyslot 0x11 was not
cleared after decrypting the ARM9 firmware binary (before jumping to the
entrypoint), and thus it was possible to gain ARM9 code execution at a later
point and instruct the AES module to regenerate all of the secret sub-keys
without having access to the decrypted keysector. This was partially fixed with
the update 9.5.0 by clearing keyslot 0x11 after ARM9Loader decrypts the ARM9
firmware binary, but keyslot 0x11 was still set with keysector Key \#1
\cite{3DS_System_Flaws}.

\subsection{Implementation (v2.0)}

The system software update 9.6.0 fixed the shortcomings of the first ARM9Loader
implementation by using a different keysector key and clearing it properly this
time. It contains the following implementation of ARM9Loader in the boot process
\cite{FIRM}:

\medskip
\begin{enumerate}
  \item Calculate SHA-256 hash of the OTP memory region and output the hash to
  the SHA\_HASH register
  \item Calculate AES write-only keyslot 0x11 from the OTP hash
  \item Read the keysector from NAND to memory
  \item Decrypt the keysector using keyslot 0x11
  \item Clear AES write-only keyslot 0x11 to zero
  \item Disable access to the OTP memory region by setting CFG\_SYSPROT9
  \item Decrypt a key from within ARM9Loader's read-only data and set that key
  to keyslot 0x18
  \item Write Key \#1 from the keysector to AES write-only keyslot 0x11
  \item Instruct the AES module to calculate sub-keys 0x19 through 0x1F based on
  keyslot 0x11
  \item Verify keyslot 0x11 by encrypting a fixed test vector and checking the
  result
  \item Write Key \#2 from the keysector to AES write-only keyslot 0x11
  \item Instruct the AES module to decrypt the ARM9 firmware binary
  \item Clear AES write-only keyslot 0x11 to zero
  \item Jump to the ARM9 firmware binary entrypoint
\end{enumerate}
\medskip

The problem with this implementation of ARM9Loader is that Key \#2 is never
verified by encrypting a fixed test vector and checking the result, meaning that
Key \#2 can be altered and the ARM9 firmware binary will be deterministically
decrypted to incorrect plaintext data and executed. Unfortunately, because the
keysector is encrypted with the console unique OTP hash (which we cannot access
post-ARM9Loader), it is not possible to arbitrarily write any key to it and get
a predictable decryption result \cite{3DS_System_Flaws}.

\section{Keyshuffling}

The keysector is encrypted with AES-128-ECB, where AES is the encryption
standard, 128 is the number of bits in a block, and ECB is the cipher mode. The
two parts of this specific encryption method that interest us are the block size
and the cipher mode. The keys in the keysector are all 16 bytes (128 bits) long,
and there is no message authentication code of any kind to increase the size or
validate the key positions. This is a crucial fact because of the cipher mode
used. In Electronic Codebook (ECB), each block in the message (the keysector in
this case) is divided into blocks of the given size and encrypted separately.
This means that each key in the keysector aligns with a block that is encrypted
completely separately from all of the other aligned keys, allowing us to move
the keys into any position we want while still decrypting properly.

Another critical aspect of this attack is that each version of the ARM9 firmware
binary is encrypted with a different counter in AES-128-CTR mode, meaning that
even code that is the same between versions will decrypt to something completely
different for each key that we try. A NAND sector on the device is 0x200 bytes
and each key is 0x10 bytes. When we do not count the Key \#2 that properly
decrypts the ARM9 firmware binary, which means we have 31 different keys for each
ARM9 firmware binary version that will all decrypt the binary to a different
incorrect plaintext which will then be executed.

\begin{figure}[h]
  \includegraphics[width=\columnwidth]{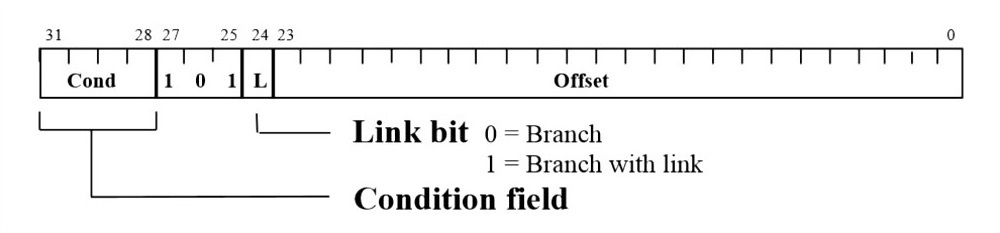}
  \caption{Encoding of an ARM branch instruction \cite{32c3}}
\end{figure}

If we try enough keys and ARM9 firmware binary versions, there is a high
probability that we will eventually find one that decrypts the ARM9 firmware
binary deterministically such that the entrypoint is a branch instruction to
another memory address where a payload can be placed. We found, by trying all
possible keys and ARM9 firmware binary versions, that there is one combination
that causes a jump to a usable memory location. By installing the 10.0.0 update
of the ARM9 firmware binary and using the keyshuffling attack to replace
keysector Key \#2 with a copy of keysector Key \#1, the resulting incorrect
plaintext from the deterministic decryption will have a jump to memory address
\texttt{0x80FD0F8} at the ARM9 firmware binary entrypoint. This redirects the
code flow outside of the secure bootchain and into manipulatable memory.

To exploit this vulnerable redirection of code flow, we took advantage of
another major oversight in the device's design: when the device reboots, all
memory keeps its contents. This makes it possible for us to gain ARM9 code
execution at a point after the system boot completes, install the 10.0.0 update
of the ARM9 firmware binary, use keyshuffling to replace keysector Key \#2 with
keysector Key \#1, insert a series of NOP instructions (``NOP sled'') at memory
address 0x80FD0F8 followed by a payload that dumps the SHA\_HASH register, then
reboot.

When the device comes back up, the ARM9 boot ROM will read ARM9Loader and the
encrypted ARM9 firmware binary to memory, then jump to ARM9Loader which will
perform the implementation v2.0 steps described previously. ARM9Loader will then
(incorrectly) attempt to decrypt the ARM9 firmware with Key \#2 (which is now
identical to Key \#1), disable access to the OTP memory region by setting
CFG\_SYSPROT9, and jump to the ARM9 firmware binary entrypoint. When it does, it
will immediately jump to memory address \texttt{0x80FD0F8}, execute the series
of NOP instructions and ``slide'' to the payload. The payload then copies the hash
of the OTP memory region from the uncleared SHA\_HASH output register for the
purpose of decrypting the keysector at a later point.

\section{Persistence}

With the SHA-256 hash of the OTP memory region, we are now able to decrypt or
re-encrypt the keysector. This means that we now completely control what key
will be used for decrypting the ARM9 firmware binary, rather than being limited
to one of the other 31 keys in the keysector. To understand how controlling the
location of memory jumped to by ARM9Loader is useful in the context of
persistence, we must look at how the boot ROM loads ARM9Loader and the firmware
binary from NAND to memory.

On the 3DS, ARM9Loader and the ARM9 Firmware binary, known collectively as
``FIRM'', are stored twice on NAND in two partitions known as ``FIRM0'' and ``FIRM1''
for redundancy purposes. This means that if one firmware partition becomes
corrupted, the device will still boot. Note that the FIRM partitions, as with
most partitions on the device, are encrypted using console unique keys derived
from the OTP and set by the boot ROM. The boot ROM uses the following
implementation to load FIRM0 and FIRM1 from NAND
\cite{clevermind}\cite{Bootloader}\cite{OTP_Registers}:

\medskip
\begin{enumerate}
  \item Decrypt the OTP memory region and store the first 0x90 bytes in
  Instruction Tightly-Coupled Memory (``ITCM'')
  \item Calculate AES write-only keyslot 0x06 from decrypted OTP memory region
  in ITCM
  \item Read FIRM0 NAND partition to memory
  \item Decrypt FIRM0 partition in memory using keyslot 0x06
  \item Check the RSA signature of decrypted FIRM0 against burned in public key
    \begin{enumerate}
    \item If the RSA signature is valid, jump to FIRM0 ARM9Loader entrypoint
    \item If the RSA signature is invalid, continue
    \end{enumerate}
  \item Read FIRM1 NAND partition to memory on top of FIRM0
  \item Decrypt FIRM1 partition in memory using keyslot 0x06
  \item Check the RSA signature of decrypted FIRM1 against burned in public key
    \begin{enumerate}
    \item If the RSA signature is valid, jump to FIRM1 ARM9Loader entrypoint
    \item If the RSA signature is invalid, panic
    \end{enumerate}
\end{enumerate}
\medskip

The problem with this implementation is that, in the case of a FIRM0 partition
with an invalid signature, FIRM1 is loaded on top of it without FIRM0's memory
being cleared \cite{32c3}. This allows for an attack in which we install the
largest legitimately signed ARM9 firmware binary available to us (8.1.0) into
FIRM0, then install the smallest legitimately signed ARM9 firmware binary
available to us (10.2.0) into FIRM1. We could then place a payload of our
choosing on top of FIRM0 at a point after FIRM1's size and find a key whose
deterministic decryption of the 10.2.0 ARM9 firmware binary to a resulting
incorrect plaintext will have a branch instruction to memory address after the
end of the 10.2.0 ARM9 firmware binary but within the size of the 8.1.0 ARM9
firmware binary \cite{32c3}.

We ran a bruteforce of all possible Key \#2 values until we found the key whose
deterministic decryption of the 8.1.0 FIRM0 ARM9 firmware binary to a resulting
incorrect plaintext has a branch instruction to 0x190 bytes after the end of the
8.1.0 ARM9 firmware binary (\texttt{0x0824D3CB4AE94D624DAA526047C59394}). We use
0x190 bytes after the end of the 8.1.0 ARM9 firmware binary because empirical
tests determined that placing the payload any sooner caused the payload to be
overwritten by an unknown factor in the boot process (likely the stack or bss
segment). After finding this key, we encrypt it with the OTP memory region hash
obtained through the keyshuffling exploit and install the encrypted key into
keysector Key \#2. We then add 0x190 to the size of the 8.1.0 ARM9 firmware
binary and write a payload of our choosing to that position relative to the
10.2.0 ARM9 firmware binary \cite{32c3}.

When the device is rebooted, the boot ROM loads FIRM0 into memory and decrypts
it with AES write-only keyslot 0x06. It then checks the RSA signature of
decrypted FIRM0, which fails because our payload at the end of the 8.1.0 ARM9
firmware binary has modified the hash. The boot ROM then, without clearing the
memory now containing our payload, loads FIRM1 into memory on top of FIRM0 and
decrypts it with AES write-only keyslot 0x06 \cite{clevermind}. The boot ROM
checks the RSA signature of FIRM1, which passes because the payload comes after
the 10.2.0 ARM9 firmware binary. The boot ROM then jumps to the FIRM1 ARM9Loader
in memory which uses our crafted Key \#2 to deterministically decrypt the 10.2.0
ARM9 firmware binary to an incorrect plaintext and jumps to its entrypoint. When
it does, it will immediately jump to the memory address of the payload of our
choosing, giving us ARM9 code execution on every successive boot before the ARM9
firmware binary runs \cite{32c3}\cite{clevermind}.

\section{Conclusion}

We have demonstrated a keyshuffling attack on the secure bootchain of the
Nintendo 3DS in order to redirect code flow into insecure memory. This allowed
us to gain code execution early enough to extract hardware secrets for the
purpose of setting up persistent early code execution that survives reboots.
This attack was made possible through a hardware revision that included the
addition of a new encryption layer that not only failed to provide extra
security, but additionally compromised a bootchain which had previously been
considered secure. This shows the danger of including new security measures in
an existing chain of trust without properly vetting them.



\bibliographystyle{IEEEtran.bst}
\bibliography{IEEEabrv,references}
%



\end{document}